\documentclass[aps,prd,onecolumn,amsmath,nofootinbib,preprintnumbers]{revtex4}

\usepackage{amsmath}
\usepackage{graphicx}
\usepackage{dcolumn}
\usepackage{bm}
\usepackage{amssymb}
\usepackage{latexsym}

\begin{document}

\title{Ricci Dark Energy in braneworld models with a Gauss-Bonnet term in the bulk}

\author{Chao-Jun Feng}
\email{fengcj@itp.ac.cn} \affiliation{Shanghai United Center for Astrophysics(SUCA), Shanghai Normal University, 100
Guilin Road, Shanghai 200234,China} \affiliation{Institute of Theoretical Physics, CAS, Beijing 100190, P.R.China}
\author{Xin-Zhou Li}
\email{kychz@shnu.edu.cn} \affiliation{Shanghai United Center for Astrophysics(SUCA), Shanghai Normal University, 100
Guilin Road, Shanghai 200234,China}


\begin{abstract}
We investigate the Ricci Dark Energy (RDE) in the braneworld models with a Gauss-Bonnet term in the Bulk. We solve the
generalized Friedmann equation on the brane analytically and find that the universe will finally enter into a pure de
Sitter spacetime in stead of the big rip that appears in the usual $4$D Ricci dark energy model with parameter $\alpha
<1/2$. We also consider the Hubble horizon as the IR cutoff in  holographic dark energy model and find it can not
accelerate the universe as in the usual case without interacting.
\end{abstract}

\pacs{}

\maketitle

The current observations such as SNeIa, CMB and SDSS et al. have strongly confirmed that our universe is accelerated
expanding. However, according to the Newton's gravity theory, ordinary matters can only attract each other, so the
universe should be decelerated expanding. Therefore, there must be something unknown that exists in the universe, and
they are often called dark energy. Experiments indicate that there are mainly about $73\%$ dark energy and $27\%$
matter components in the recent universe, but  people still do not understand what is dark energy from fundamental
theory currently.

The cosmology constant which can only appears in the Einstein equation without ruining the Bianch identity seems the
best candidate for dark energy, but it suffers from the fine-tuning and coincident problems. The fine-tuning problem
says why the vacuum energy density observed today is so much smaller than the value predicted by the quantum field
theory, and we need to fine tune the cosmological constant to cancel it and get the observation value, while the
coincidence problem is asking why it is today that dark energy becomes important, since its evolution behavior is so
much different from that of ordinary matter. In order to  alleviate these problems, a lot of dynamic dark energy models
have been built, such as quintessence, phantom, quintom models which are basically scalar field models. Another way to
explain the accelerating is to modify general gravity like $f(R)$ theory and DGP model, etc..

Actually, the cosmology constant (or dark energy) problem is in essence an issue of quantum gravity
\cite{Witten:2000zk}, since the energy density of dark energy is inevitably related to the large vacuum energy density
of the quantum field theory without including gravity. Considering the gravity effects, there may be some regions in
which the field theory is invalid. The holographic principle regards the black hole as the object with maximum entropy
in a given region, and from statistical physics, the entropy is a extensive quantity which proportional to the volume
of such region, while the black hole' entropy is proportional to the area of its surface, so in the field theory, there
should exists a infrared (IR) cutoff, beyond which the field theory will be failed. However, such constraint seems a
little bit loose, because it includes the black hole state in the field theory. To avoid the existence of such states,
Cohen et al.\cite{CohenZX:1999} suggested that in a given region with length scale $L$, the field's energy should be
bounded by the black hole's, i.e. $\rho L^3 \le LM^2_{pl}$, where $\rho$ is the total energy density within the region
and $M_{pl} = G^{-1/2}$ is the Planck mass. Applying the holographic principle to cosmology, Li \cite{Li:2004rb} has
proposed the holographic dark energy model, in which the  energy density of dark energy is $\rho = 3c^2M^2_{pl}L^{-2}$,
namely it saturates the bound. He finds that when $L = R_h$, which is the future event horizon, this model will be
consistent with observations and meanwhile solves the coincidence problem, and this model is often called HDE for
short.

Although the holographic model based on the future event horizon is successful in fitting the current data, some
authors asked why the current acceleration of the universe is determined by its future. Actually, the future event
horizon is not the only choice for the holographic dark energy model. Also motivated by the holographic principle, Gao,
et al.\cite{Gao:2007ep} have proposed the Ricci dark energy (RDE) model recently, in which the future event horizon
area is replaced by the inverse of Ricci scalar, and this model is also phenomenologically viable.

Assuming the black hole is formed by gravitation collapsing of the perturbation in the universe, the maximal black hole
can be formed is determined by the casual connection scale $R_{CC}$ given by the "Jeans" scale of the perturbations.
For tensor perturbations, i.e. gravitational perturbations, $R_{CC}^{-2} = Max(\dot H + 2H^2, -\dot H)$ for a flat
universe, and according to the ref.\cite{Cai:2008nk}, only in the case of $R_{CC}^{-2} =\dot H + 2H^2$, it could be
consistent with the current cosmological observations when the vacuum density appears as an independently conserved
energy component. Therefore, if one choices the casual connection scale $R_{CC}$ as the IR cutoff in the holographic
dark energy, the Ricci dark energy model is also obtained. For recent progress on Ricci dark energy and holographic
dark energy, see ref.\cite{recentRicci}\cite{Gra}\cite{holo:recent}\cite{brane GB}\cite{Zhang:constraint}.

In flat FRW universe, the Ricci scalar is $R_4=6(\dot H + 2H^2)$ and the energy density of RDE reads
\begin{equation}\label{energy density}
    \rho_{R} = 3\alpha \left(\dot H + 2H^2 \right) \propto R_4\,,
\end{equation}
where we have set $8\pi G = 1$ and  $H = \dot a / a$ is the Hubble parameter. Here $\alpha$ is a dimensionless
parameter which will determine the evolution behavior of RDE.

In ref.\cite{brane GB}, the author apply bulk holographic principle in general branewold models with a Gauss-Bonnet
term in the bulk, and get an effective $4$D holographic dark energy from $5$D theory. They have also taken the $4$D
future event horizon as the IR cutoff to study the behaviors of the dark energy. In this letter, we investigate the RDE
model in this braneworld model and find that the RDE model which behaviors like a quintom with parameter $\alpha<1/2$
and dominates the universe in the future will lead the universe to enter into the pure de Sitter spacetime instead of
the big rip in the far future. Early works on dark energy models with late time de Sitter attractor, see
ref.\cite{XinzhouLi}.

The $5$D braneworld models with a Gauss-Bonnet term in the bulk can be described by the following action \cite{brane
GB}:
\begin{eqnarray}
  \nonumber
  S = &\int& d^4x dy \sqrt{-g}(M_5^3R - \rho_{\Lambda5} + \beta M_5 \mathcal{L}_{GB}) \\
      &+& \int d^4x \sqrt{-\gamma}(\mathcal{L}_{br}^{mat} - V + r_cM^3_5 R_4)\label{action} \,,
\end{eqnarray}
where $M_5$ is the $5$D Planck mass, $\rho_{\Lambda5}$ is the bulk cosmological constant, and $R$ is the curvature
scalar of the $5$D bulk spacetime with metric $g_{AB}$. As usual, the Gauss-Bonnet lagrangian is
\begin{equation}\label{gb lag}
    \mathcal{L}_{GB} = R^2 - 4R_{AB}R^{AB} + R_{ABCD}R^{ABCD}
\end{equation}
with coupling constant $\beta$, and $R_{ABCD}$, $R_{AB}$ are the Riemann tensor and Ricci tensor respectively. In the
second intergral, $\gamma = \det g_{\alpha\beta}$ is the induced $4$D metric, $V$ is the brane tension and
$\mathcal{\L}_{br}^{mat}$ is an arbitrary brane matter content. The last term is arisen from radiative corrections,
with $r_c$ its characteristic length scale and $R_4$ the $4$D Ricci scalar. By using the Gaussian normal coordinates
with the metric:
\begin{equation}\label{metric}
    ds^2 = -n^2(\tau, y)d\tau^2 + a^2(\tau,y)d\Omega_k^2 + dy^2 \,,
\end{equation}
we can imposed a $Z_2$-symmetry around the brane located at $y = 0$ , i.e. $n(\tau, y =0) = 1$ to get a cosmology
evolution on the brane. Here $d\Omega_k^2$ denotes the metric in a maximally symmetric $3$-dimensional space with
$k=-1,0,+1$ parameterizing its spacial curvature. Matters on the brane are regarded as perfect fluids with equation
state $p = w\rho$. After integration of the $00$ and $ii$ components of the $5D$ Einstein equations around the brane,
we get the brane cosmology evolution equations in the low-energy limit $\rho \ll V$ as in \cite{brane GB} :
\begin{widetext}
\begin{equation}\label{Friedmann}
    H^2 + \frac{k}{a^2} = (72M_5^6 - 16\beta\rho_{\Lambda5}M_5^3 + 6r_c V M_5^3)^{-1} V\rho
   + \frac{V^2}{144M_5^6} - \frac{1-\sqrt{1+\tilde\Lambda}}{36\beta}\bigg(2+\sqrt{1+\tilde\Lambda}\bigg)^2
\end{equation}
\end{widetext}
where $\tilde\Lambda = 2\beta\rho_{\Lambda5}/(3M_5^3)$. Here $a$ is the scale factor on the brane. In order to be
consistent with conventional $4$D Friedmann equation, one has to impose:
\begin{equation}
    \quad V = \frac{192\pi M_5^6}{M_p^2 - 16\pi r_cM_5^3} \,.
\end{equation}
Then the evolution equation (\ref{Friedmann})becomes:
\begin{equation}\label{Friedmann2}
    H^2 + \frac{k}{a^2} = \frac{8\pi \rho}{3M_p^2}\bigg(1+V_1(\beta, \rho_{\Lambda5})\bigg)  + \frac{8\pi}{3M_p^2}\rho_\Lambda
    \,,
\end{equation}
where $V_1$ is defined as
\begin{equation}\label{V1}
    V_1(\beta, \rho_{\Lambda5}) \equiv \frac{128\pi\beta\rho_{\Lambda5}}{3VL_5-128\pi\beta\rho_{\Lambda5}}
    \,,
\end{equation}
and here we have used the relation between the $5$D and $4$D Planck mass $M_5^3 = M_p^2L_5^{-1}$, where $L_5$ is the
size of the extra dimension. The $4$D effective dark energy  $\rho_\Lambda \equiv \rho_{\Lambda4}^{eff}$ in
eq.(\ref{Friedmann2}) is:
\begin{equation}\label{4d rho}
    \rho_\Lambda  = \frac{96\pi M_p^2}{(L_5-16\pi r_c)^2} - \frac{M_p^2}{96\pi
    \beta}\left(1-\sqrt{1+\tilde{\Lambda}}\right)\left(2+\sqrt{1+\tilde{\Lambda}}\right)^2
\end{equation}

A $D$-dimensional spherical and uncharged black hole with mass $M_{BH}$ as
\begin{equation}\label{mass of bh}
    M_{BH} = r_s^{D-3}(\pi)^{\frac{D-3}{2}} (M_D)^{D-2}\frac{D-2}{8\Gamma(\frac{D-1}{2})}\,,
\end{equation}
where $r_s$ is its Schwarzschild radius and $M_D$ is D-dimensional Planck mass, namely $(M_D)^{D-2} = G_D^{-1}$ and it
relates to the usual $4$-dimensional Planck mass $M_p$ by $M_p^2 = (M_D)^{D-2}V_{D-4}$, where $V_{D-4}$ is the volume
of the extra-dimensional space. Let $\rho_{\Lambda D}$ is the vacuum energy in the bulk and by applying $D$-dimensional
holographic principle we obtain:
\begin{equation}\label{holo}
    \rho_{\Lambda D} Vol(S^{D-2}) \leq r^{D-3}(\pi)^{\frac{D-3}{2}} (M_D)^{D-2}\frac{D-2}{8\Gamma(\frac{D-1}{2})}\,,
\end{equation}
where $Vol(S^{D-2})$ is the volume of the maximal hypersphere in a $D$-dimensional spacetime, and it reads
$Vol(S^{D-2}) = A_Dr^{D-1}$ with
\begin{equation}
    A_D = \frac{\pi^{\frac{D-1}{2}}}{ \left(\frac{D-1}{2}\right)! }  \,,\quad
    A_D = \frac{\left(\frac{D-2}{2}\right)!}{(D-1)!} 2^{D-1}\pi^{\frac{D-2}{2}}\,,
\end{equation}
for $D-1$ being even or odd respectively. Therefore, by saturating inequality(\ref{holo}), we get the $D$-dimensional
holographic dark energy as
\begin{equation}\label{holo de}
   \rho_{\Lambda D} = c^2(\pi)^{\frac{D-3}{2}} (M_D)^{D-2} A_D^{-1}\frac{D-2}{8\Gamma(\frac{D-1}{2})} L^{-2}\,.
\end{equation}
Therefore, the $5$D holographic dark energy is given by:
\begin{equation}\label{5d holo}
    \rho_{\Lambda5} = 3c^2 \frac{1}{4\pi}M^3_5 L^{-2} \,.
\end{equation}
Hereafter, we will use the assumption in ref.\cite{brane GB} that $L_5$ is arbitrary large finitely, namely, it is
larger than any other length of this model leaving the brane evolution independent of the size of the bulk and this is
the reason for the single-brane consideration. So,we will use the approximation $L_5\gg r_c$ and $L_5^{-1} \rightarrow
0$ in the calculation . Furthermore, we also expand relevant quantities in terms of the Gauss-Bonnet coupling $\beta$,
and only keep linear terms, thus we obtain
\begin{equation}
    V_1(\beta, L) = \beta\frac{c^2}{6\pi}L^{-2} + \mathcal{O}(\beta^2)\,,
\end{equation}
and the effective $4$D energy density for the dark energy from (\ref{4d rho}):
\begin{equation}\label{4d effective}
    \rho_\Lambda \equiv \rho_{\Lambda4}^{eff} = 3c^2\frac{1}{128\pi^2}M_p^2L^{-2}\left(1+\beta\frac{c^2}{24\pi}L^{-2}\right)  + \mathcal{O}(\beta^2)
    \,.
\end{equation}
Comparing with the result directly derived from eq.(\ref{holo de}):
\begin{equation}\label{4d direct}
    \rho_{\Lambda4} = \frac{3c^2}{8\pi}M_p^2L^{-2} \,,
\end{equation}
where one can see the effective dark energy $\rho_{\Lambda4}^{eff}$ is smaller than $\rho_{\Lambda4}$ for the same IR
cutoff $L$, which means the dark energy derived from $5$D holographic principle is stringent than the one directly from
$4$D. Therefore, we get the Friedmann equation for flat brane universe under these approximation:
\begin{equation}\label{Friedmann3}
     3H^2 = \rho\bigg(1+ \frac{8}{3}\beta\alpha  L^{-2} \bigg) + 3\alpha L^{-2} \bigg(1+ \frac{2}{3}\beta\alpha  L^{-2} \bigg)
    \,,
\end{equation}
where we have set $8\pi M_P^{-2} = 1$ and $\alpha \equiv c^2/(16\pi)$ for convenience. When $\beta\rightarrow0$, the
above equation return to the result in ref.\cite{Gao:2007ep}. Here one can see, there is interacting terms between dark
energy and ordinary matter, and it arises naturally from the $5$D dynamics and the use of holographic principle in the
bulk, rather than putting by hand like interacting dark energy. The IR cutoff $L$ could be Hubble horizon, particle
horizon or future event horizon etc. and there is no first principle to judge which cutoff is correct. Therefore, the
only way to check these IR cutoff is by the observations. As we mentioned above, once the $5$D holographic principle is
satisfied, it should be also satisfied in $4$D, because the derived effective dark energy (\ref{4d effective}) from
$5$D bulk is stringed than that(\ref{4d direct}) from $4$D itself. Therefore, we would like to identify the IR cutoff
$L$ to the "Jeans-like" causal connection length scale $R_{CC}$, beyond which the black hole can not formed by
gravitational collapse of perturbation in the universe. For gravitational perturbation and not conflict with
experiments, $R_{CC}^{-2} = \dot H + 2H^2$ is the only choice for $4$D flat universe. By taking the $R_{CC}$ as the IR
cutoff in eq.(\ref{4d direct}), the Ricci dark energy model is obtained, but here we will using the effective $4$D dark
energy (\ref{4d effective}) to study the Ricci dark energy in the braneworld models.

Before investigating the Ricci dark energy in this model, we would like to taking a simple choice $L = H^{-1}$, and in
this case, we shall see that it can not make the universe accelerating.  The Friedmann equation (\ref{Friedmann3}) can
be rearranged:
\begin{equation}
    2\alpha\beta\left(\alpha H^2\right)^2 + \left(3\alpha + \frac{8}{3}\alpha\beta\rho-3\right)\alpha H^2 + \alpha \rho = 0 \,,
\end{equation}
and the solution to this equation is the following:
\begin{equation}\label{sol to lh}
    H^2 = \frac{\rho}{3(1-\alpha)} + \frac{2\alpha(4-3\alpha)\rho^2}{27(1-\alpha)^3 } \beta \,,
\end{equation}
where we have kept the linear term of $\beta$ only. Applying the energy conservation equation and assuming only matters
(pressureless) resides on the brane universe, we get $\rho_m = \rho_{m0}(a/a_0)^{-3}$, where the subscript $0$ denotes
the present value of the quantity. Therefore, we get the evolution of the universe on the brane:
\begin{equation}\label{sol to lh a}
    a(\tau) = a_0\left( C_0 \tau^{\frac{2}{3}} + C_1 \beta \tau^{-\frac{4}{3}}\right) \,,
\end{equation}
where
\begin{equation}
    C_0 = \left[\frac{3\rho_{m0}}{4(1-\alpha)}\right]^{\frac{1}{3}} \,, \quad C_1 = \frac{2^{\frac{7}{3}}}{3^{\frac{11}{3}}}
    \frac{(4-3\alpha)\alpha}{(1-\alpha)^{\frac{4}{3}}}\rho_{m0}^{\frac{1}{3}} \,.
\end{equation}
Thus, one can see the first term on the rhs. of eq.(\ref{sol to lh a}) is much like the usual evolution of matter
dominated universe, while the second one is the correction from the Gauss-Bonnet term. The energy density of the
effective $4$D dark energy is
\begin{equation}
    \rho_\Lambda = \frac{\alpha\rho}{1-\alpha} + \frac{2\alpha^2(5-4\alpha)\rho^2}{9(1-\alpha)^3}\beta \,,
\end{equation}
and the its equation of state parameter is
\begin{equation}
    w \equiv -1-\frac{1}{3}\frac{d\ln\rho_\Lambda}{d\ln a} = \frac{2\alpha(5-4\alpha)\rho_{m0}}{9(1-\alpha)^2}\beta
    (1+z)^3 >0 \,.
\end{equation}
So, the universe is not accelerated expanding and $L=H^{-1}$ is not a good choice for IR cutoff in this sense.

\begin{figure}[h]
\includegraphics[width=0.4\textwidth]{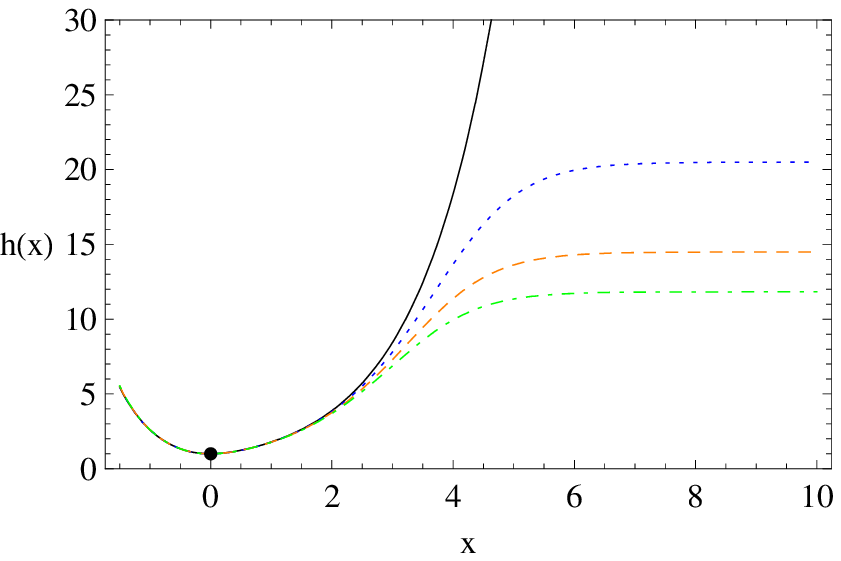}
\qquad
\includegraphics[width=0.4\textwidth]{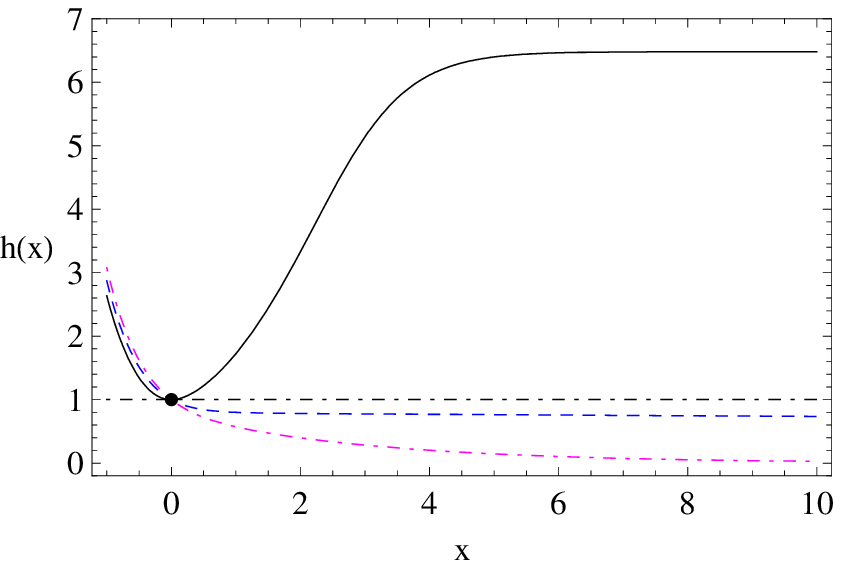}
\caption{\label{fig::hubble}Evolution of h(x) with initial condition h(0) = 1 and $\Omega_{m0} = 0.27$. The left figure
is plotted with $\alpha = 0.36$ and $\tilde\beta =0$(solid curve), $0.001$(dotted curve), $0.002$(dashed curve),
$0.003$(dotdashed curve), and it shows the Hubble parameter will blow up in sometime without the Gauss-Bonnet term
($\tilde\beta = 0$) while the universe will finally enter into the de Sitter space with the help of the Gauss-Bonnet
term ($\tilde\beta \neq 0$). The right figure indicates is plotted with $\tilde\beta = 0.01$ and $\alpha = 0.36$(solid
curve), $0.5$(dashed curve), $0.6$(dotdashed curve), while the horizon dotdashed line corresponds $h(x) = -1$. }
\end{figure}

Now, we will identify $L^{-2}=R_{CC}^{-2}=\dot H + 2H^2$ in flat universe, then the Friedmann equation
(\ref{Friedmann3}) becomes:
\begin{equation}
    \bigg(2\beta\alpha^2 H^2\bigg) Y^2 + \left(\frac{8}{3}\rho\beta\alpha  +  3\alpha\right) Y + \rho H^{-2} - 3 = 0
    \,,
\end{equation}
where we have defined $Y\equiv (R_{CC}H)^{-2}$, and by solving this equation we get:
\begin{equation}\label{Y equ}
    Y = \frac{3-\rho H^{-2}}{3\alpha} - \frac{6H^2 + 4\rho - 2\rho^2 H^{-2}}{9\alpha} \beta \,,
\end{equation}
up to the linear term of $\beta$. Replacing $Y H^{2}= (H^2)'/2 + 2H^2$ and $\rho = \rho_{m0}e^{-3x}$ we get the
differential equation for the Hubble parameter as follows:
\begin{widetext}
\begin{equation}\label{Friedmann4}
    \frac{(h^2)'}{2} + \left(2-\frac{1}{\alpha}\right)h^2 + \frac{\Omega_{m0}e^{-3x}}{\alpha} + \frac{2h^4 + 4h^2\Omega_{m0}e^{-3x}  - 6(\Omega_{m0}e^{-3x})^2 }{3\alpha} \tilde\beta  = 0\,,
\end{equation}
\end{widetext}
where $\Omega_{m0} = \rho_{m0}/(3H^2_0)$ and we have defined dimensionless coupling $\tilde\beta = H_0^2\beta$.
Hereafter, prime denotes the derivative with respect to $x\equiv \ln a $. Substituting
\begin{equation}
    u(x) = \exp{\left[\frac{4\tilde\beta}{3\alpha} \int h^2 dx + \left(4-\frac{2}{\alpha}\right)x \right]}
\end{equation}
reduce the eq. (\ref{Friedmann4}) to a second-order linear equation of $u$:
\begin{widetext}
\begin{equation}
    9\xi\frac{d^2u}{d\xi^2} + \left(-\frac{8\Omega_{m0}\tilde\beta}{\alpha}\xi + 21 - \frac{6}{\alpha}\right) \frac{du}{d\xi} +
    \frac{8\Omega_{m0}\tilde\beta}{3\alpha}\left(- \frac{2\Omega_{m0}\tilde\beta}{\alpha}\xi - 4 +\frac{3}{\alpha} \right) u =
    0 \,,
\end{equation}
\end{widetext}
where $\xi = e^{-3x}$ and the solution to this equation is
\begin{equation}
    u(\xi) = {}_1F_1\left(\frac{13}{12}, \, \frac{7}{3} - \frac{2}{3\alpha}, \,-\frac{16\Omega_{m0}\tilde\beta}{9\alpha}\xi\right)
    \exp{\left(\frac{4\Omega_{m0}\tilde\beta}{3\alpha}\xi\right)} \,,
\end{equation}
where ${}_1F_1$ is the Kummer confluent hypergeometric function. When $x\rightarrow \infty$, i.e. $\xi \rightarrow 0$,
the Kummer function $ {}_1F_1 \approx 1$, and also $u(\xi)\approx 1$, thus we obtain the final value of the Hubble
parameter
\begin{equation}
    h(x\rightarrow\infty) \approx \sqrt{\frac{3\left(1-2\alpha\right)}{2\tilde\beta}}\,.
\end{equation}
Therefore, if $\tilde\beta = 0$, which means no Gauss-Bonnet term in the action (\ref{action}), the Hubble parameter
will become infinite in the future. And it is due to the Gauss-Bonnet term that the big rip disappears and finally the
energy density of the dark energy is
\begin{equation}
    \rho_\Lambda(x\rightarrow\infty) \approx 3\alpha\bigg(1-8\alpha^2+8\alpha^3\bigg)\frac{\rho_c}{\tilde\beta} \,,
\end{equation}
where $\rho_c \equiv 3H_0^2$ is the critical energy density today. According the holographic principle, the UV cutoff
satisfy $\Lambda^2 \leq L^{-1}M_p$ , and the inverse of IR cutoff $L^{-1}$ should be smaller than the UV cutoff by
definition, i.e. $L^{-1} < \Lambda$, so $\Lambda < M_p$. Therefore, it requires
\begin{equation}
    \tilde\beta \gtrsim \frac{\rho_c}{M_p^4} \sim 10^{-120} \quad \text{or} \quad  \beta \gtrsim l_p^2 \,,
\end{equation}
where $l_p$ is the Planck length. To illustrate and double check our analytical solution, we also give an example of
numerical calculation in Fig.\ref{fig::hubble},
from which one can see the effect of the Gauss-Bonnet term as follows: when $\tilde\beta = 0$, we recover the usual
Ricci dark energy in ref.\cite{Gao:2007ep}, then it shows the Hubble parameter will blow up in finite time, which means
there will be a big rip in some time. However, the actually framework of the holographic model of dark energy is the
effective field theory with an UV/IR duality. Before the big rip, the UV cutoff of the theory will exceed the Planck
energy scale, and this will definitely spoil the effective field theory. Therefore, in the holographic model of dark
energy, the big rip is actually not allowed. Now, with the help of the Gauss-Bonnet term, the universe will finally
enter into the steady state (de Sitter) space instead of the big rip, see Fig.\ref{fig::hubble}. If the parameter
$\alpha \ge 1/2$, there is no big rip time in the later time with or without the Gauss-Bonnet term, see the right one
in Fig.\ref{fig::hubble}.

In ref.\cite{Gra}, the author generalized the IR cutoff for holographic dark energy as $L^{-2} \propto \gamma H^2 +
\lambda\dot H$, where $\gamma$ and $\lambda$ are constants. When $\gamma\lambda^{-1} = 2$, it reduces to RDE. By taking
$Y = \dot H H^{-2} + \gamma\lambda^{-1}$ and from eq.(\ref{Y equ}), we obtain
\begin{equation}\label{Friedmann gen}
    \frac{(h^2)'}{2} + \frac{(\gamma - 1)}{\lambda}h^2 + \frac{\Omega_{m0}e^{-3x}}{\lambda} + \frac{2h^4 + 4h^2\Omega_{m0}e^{-3x}  - 6(\Omega_{m0}e^{-3x})^2 }{3\lambda} \tilde\beta  =
    0\,.
\end{equation}
Substituting
\begin{equation}
    v(x) = \exp{\left[\frac{4\tilde\beta}{3\lambda} \int h^2 dx +\frac{2(\gamma - 1)}{\lambda} x \right]}
\end{equation}
reduce the eq. (\ref{Friedmann gen}) to a second-order linear equation of $v$:
\begin{widetext}
\begin{equation}
    9\xi\frac{d^2v}{d\xi^2} + \left(-\frac{8\Omega_{m0}\tilde\beta}{\lambda}\xi + 9 + \frac{6(\gamma - 1)}{\lambda}\right) \frac{du}{d\xi} +
    \frac{8\Omega_{m0}\tilde\beta}{3\lambda}\left(- \frac{2\Omega_{m0}\tilde\beta}{\lambda}\xi + \frac{3-2\gamma}{\lambda} \right) v =
    0 \,,
\end{equation}
\end{widetext}
and its solution is
\begin{equation}
    v(\xi) = {}_1F_1\left(\frac{3}{4}+\frac{\gamma}{6\lambda}, \, 1 + \frac{2(\gamma -1)}{3\lambda}, \,-\frac{16\Omega_{m0}\tilde\beta}{9\lambda}\xi\right)
    \exp{\left(\frac{4\Omega_{m0}\tilde\beta}{3\lambda}\xi\right)} \,,
\end{equation}
and we also get the finally value of Hubble parameter:
\begin{equation}
   h(x\rightarrow\infty) \approx \sqrt{\frac{3\left(1-\gamma\right)}{2\tilde\beta}}\,.
\end{equation}
Again, the universe will enter into de Sitter spacetime instead of the big rip in the far future. Then, the energy
density of dark energy at that time will be
\begin{equation}
    \rho_\Lambda(x\rightarrow\infty) \approx \frac{3\gamma}{2}\bigg(1-2\gamma^2+\gamma^3\bigg)\frac{\rho_c}{\tilde\beta}\,,
\end{equation}
and it also requires $\tilde\beta \gtrsim \rho_cM_p^{-4}$ or $\beta \gtrsim l_p^2$.

In conclusion, we have investigate the Ricci dark energy in the framework of $5$D braneworld models with a Gauss-Bonnet
term in the bulk. Due to the effect of the Gauss-Bonnet term, the universe will avoid the big rip problem that appears
in the usual Ricci dark energy with parameter $\alpha <1/2$ . In that case, the universe will finally dominated by the
Ricci dark energy and the scale factor will becomes infinity in finite time, while in our result, the universe will
enter into the de Sitter space in the far future instead of the big rip. Since the holographic Ricci dark energy is
basically the effective field theory with UV/IR duality, the big rip seems inconsistent with the theoretical framework,
therefore the story of Ricci dark energy in the braneworld model with a Gauss-Bonnet term sounds interesting and is
deserved to be studied further.

\begin{acknowledgments}
This work is supported by National Science Foundation of China grant No. 10847153 and No. 10671128.
\end{acknowledgments}

\end{document}